\documentclass[aps,preprint,showpacs,preprintnumbers,amsmath,amssymb]{revtex4}
\usepackage{amsmath,mathrsfs,amsbsy,color,graphicx,bm,amsthm,amsfonts}
\usepackage{units}
\usepackage{bbm}
\usepackage{times}
\usepackage{dcolumn}
\usepackage{mathrsfs}
\usepackage{amsmath,amssymb,epsfig}
\usepackage{amsmath}
\newcommand{\udots}{\mathinner{\mskip1mu\raise1pt\vbox{\kern7pt\hbox{.}}
\mskip2mu\raise4pt\hbox{.}\mskip2mu\raise7pt\hbox{.}\mskip1mu}}
\begin{document}

\title{ Fermionic steering and its  monogamy relations in Schwarzschild spacetime  }
\author{Shu-Min Wu$^1$, Hao-Sheng Zeng$^2$\footnote{Email: hszeng@hunnu.edu.cn}}
\affiliation{$^1$ Department of Physics, Liaoning Normal University, Dalian 116029, China\\
$^2$ Department of Physics, Hunan Normal University, Changsha 410081, China}


\begin{abstract}
Using two different types of quantification for quantum steering, we study the influence of Hawking radiation on quantum steering for fermionic fields in Schwarzschild spacetime. The degradation for the steering between physically accessible observers and the generation for the steering between physically accessible and inaccessible observers induced by Hawking radiation are studied. We also reveal the difference between the two types of quantification for steering, and find some monogamy relations between steering and entanglement. Furthermore,  we show the different properties between fermionic steering and bosonic steering in Schwarzschild spacetime.

\end{abstract}

\vspace*{0.5cm}
\maketitle
\section{Introduction}
Einstein-Podolsky-Rosen steering, firstly introduced by Schr\"{o}dinger in 1935, is an intermediate type of quantum correlation between Bell non-locality and quantum entanglement \cite{Q1,Q2}.
Quantum steering is a special quantum phenomenon, which allows an observer by measuring his subsystem to control the remote subsystem owned by another observer.
Since Wiseman $et al.$ in 2007 firstly gave the resource definition of quantum steering via  the local hidden state (LHS) model \cite{Q3}, quantum steering has attracted renewed interest \cite{Q4,Q5,Q6,Q7,Q8,Q9,Q10,Q11,Q12,Q13}. Various inequalities for detecting quantum steering of quantum states have been proposed \cite{Q14,Q15,Q16,Q17,Q18,Q19}.
Unlike quantum entanglement, the asymmetry of quantum steering in quantum systems, such as one-way  steering and two-way  steering, has been demonstrated experimentally \cite{Q20,Q21,Q22,Q23}.
One-way  steering and two-way  steering are very important quantum resources, which may be used in quantum information processing \cite{Q24,Q25,Q26}.

General relativity from the Einstein's theory has made great achievements and its predictions have been gradually verified in physics and astronomy. One of the most famous predictions is that  black holes exist in our universe.  Black holes may be created by
the gravitational collapse of sufficiently massive stars.
Recently,  the Advanced LIGO and Virgo detectors detected  the gravitational wave (GW150914) from a binary black hole merger system for the first time, which indirectly
confirms the existence of black holes and Einstein's general relativity \cite{Q27}.
Moreover, the first photo of the supermassive black hole in the center of the giant elliptical galaxy M87 was taken by the Event Horizon Telescope \cite{Q28,Q29,Q30,Q31,Q32,Q33}.
Soon after,  Holoien $et al.$ shown that a star gets torn apart by a black hole \cite{Q34}. Because black holes are too far away from us and have special properties, it is covered with a mysterious veil and is the cutting-edge research field, such as the lost information paradox of black holes \cite{Q35,Q36,Q37}.
Hawking predicted that the vacuum fluctuations near the event horizon cause black holes to evaporate. In other words, particle-antiparticle pairs near the event horizon are generated. Hawking speculated that  antiparticle falls into the black hole and particle escapes from the black hole.
Obviously, the Hawking radiation underlies the lost information paradox of black holes.

The combination of relativity theory and another fundamental
theory of modern physics, quantum information, gave birth to relativistic quantum information.
It is believed that the study of quantum information in a
relativistic framework is not only helpful in understanding
concepts of quantum correlation and coherence \cite{Q38,Q39,Q40,Q41,Q42,Q43,Q44,Q45,Q46,Q47,Q48,Q49,Q50,Q51,Q52,Dong1,Dong2,Dong3,Dong4,Dong5,Dong6,Dong7,MLO,MLO1}, but also plays an important role in the investigation of the lost information paradox of black holes \cite{Q35,Q36,Q37}.
It has been already shown that in the relativistic quantum information, the bosonic and fermionic fields have different behaviors. For example, in the limit of infinite acceleration, bosonic entanglement vanishes \cite{Q38}, while fermionic entanglement can survive \cite{Q39}; With the increase of acceleration, bosonic discord appears irreversible decoherence \cite{Q46}, while fermionic discord occurs the phenomenon of freeze \cite{Q47}.
Compared with quantum entanglement, quantum steering has richer properties, such as no-way  steering,  one-way  steering, two-way  steering and asymmetric steering. Therefore, studying bosonic and fermionic steering, and comparing them in the relativistic framework are more intriguing.

In this work, we will study the influence of Hawking radiation on quantum steering for fermionic fields in Schwarzschild spacetime. We assume that Alice and Bob share a maximally entangled state of Dirac fields  in flat Minkowski spacetime. Afterwards, Alice continues to stay at an asymptotically flat region, while Bob hovers near the event horizon of the black hole; at the same time, the Anti-Bob inside the event horizon is
generated. We will calculate fermionic steering and obtain the analytic expressions in the curved spacetime.
We then study the degradation of steering between Alice and Bob, the production of steering between Alice and Anti-Bob, Bob and Anti-Bob. We also study the redistribution of steering between different subsystems and try to find some monogamy relations for steering and entanglement.

The paper is organized as follows. In Sec.II, we briefly introduce the quantification of bipartite steering. In Sec.III, we discuss the quantization of Dirac fields in the background of a Schwarzschild black hole. In Sec.IV, we study the evolution of steering between different subsystems and their redistribution rules, and make a comparison with the counterpart of bosonic fields in the background of a Schwarzschild black hole. In Sec.IV, we study the monogamy relations between fermionic steering and entanglement in Schwarzschild spacetime. Finally the summary is arranged in Sec.VI.

\section{Quantification of bipartite steering }

\subsection{Quantification of steering based on entropy uncertainty relation }
According to the definition of quantum steering given by Wiseman $et al.$, one can say that Alice can steer Bob's state if the results presented by Alice and Bob have correlations that violate the local-hidden-state (LHS) model \cite{Q3}. Consider a state $\rho_{AB}$ of the discrete-variable compound system $AB$, with subsystems $A$ and $B$ held by Alice and Bob respectively. Let $\hat{R}^A$ and $\hat{R}^B$ are the discrete observables for subsystems $A$ and $B$, with possible outcomes $\{R^A\}$ and $\{R^B\}$. The LHS model may be written as \cite{Q14}
\begin{eqnarray}\label{w1}
H(R^B|R^A)\geq \sum_\lambda P(\lambda)H_Q(R^B|\lambda),
\end{eqnarray}
where $H(R^B|R^A)$ is the conditional entropy of variable $\hat{R}^B$ given $\hat{R}^A$ to be $R^A$, and $H_Q(R^B|\lambda)$ is the discrete Shannon entropy of the probability distribution $P_Q(R^B|\lambda)$ that measures $\hat{R}^B$ to be $R^B$ given the details of preparation in the hidden variable $\lambda$.

In the $N$-dimensional Hilbert space, any pair of discrete observables $R$ and $S$  with the eigenbases $R_i$ and $S_i$ ($i=1,2,..,N$)  satisfy the entropy uncertainty relation
\begin{eqnarray}\label{w2}
H_Q(R)+H_Q(S)\geq \log(\mathrm{U}),
\end{eqnarray}
where $\mathrm{U}\equiv{\rm{min}}_{i,j}\frac{1}{|\langle R_i|S_j\rangle|^2}$.
From above two equations, one can obtain the entropy-based steering inequality
\begin{eqnarray}\label{w3}
H(R^B|R^A)+H(S^B|S^A)\geq \log(\mathrm{U}^B),
\end{eqnarray}
where $\mathrm{U}^B$ denotes the value of $\mathrm{U}$ associated with observables $R^B$ and $S^B$.
This steering inequality that involves a pair of discrete observables may be generalized to the more general case that involves arbitrary number of mutually unbiased observables. Especially for the complete set of pairwise complementary Pauli operators $X$, $Y$ and $Z$, there exists the steering inequality from Alice to Bob \cite{Q14}
\begin{eqnarray}\label{w4}
I^{A\rightarrow B}=H(\sigma_x^B|\sigma_x^A)+H(\sigma_y^B|\sigma_y^A)+H(\sigma_z^B|\sigma_z^A)\geq2.
\end{eqnarray}
If this steering equality is violated, we say that Alice can steer Bob. To quantify the ability that Alice steers Bob, one introduce the quantity
\begin{eqnarray}\label{w5}
S^{A\rightarrow B}={\rm{max}}\bigg\{0, \frac{I^{A\rightarrow B}-2}{I_{\rm{max}}-2}\bigg\}.
\end{eqnarray}
The factor $I_{\rm{max}}$ is to guarantee the normalization of quantum steerability, which is equal to 6 for the maximally entangled state considered in our text.

In this paper, we study symmetric X-state
\begin{eqnarray}\label{w6}
\rho_x= \left(\!\!\begin{array}{cccc}
\rho_{11}&0&0&\rho_{14}\\
0&\rho_{22}&\rho_{23}&0\\
0&\rho_{23}&\rho_{33}&0\\
\rho_{14}&0&0&\rho_{44}
\end{array}\!\!\right),
\end{eqnarray}
where the real entries satisfy $\rho_{ij}=\rho_{ji}$.
This X-state can also be expressed in the form
\begin{eqnarray}\label{w7}
\rho_x=\frac{1}{4}\bigg[I\otimes I+p\sigma_z\otimes I+qI\otimes\sigma_z+\sum_{i=1}^3c_i\sigma_i\otimes\sigma_i \bigg],
\end{eqnarray}
where $c_1=2(\rho_{14}+\rho_{23})$,  $c_2=2(\rho_{23}-\rho_{14})$,  $c_3=\rho_{11}-\rho_{22}-\rho_{33}+\rho_{44}$, $p=\rho_{11}+\rho_{22}-\rho_{33}-\rho_{44}$
and $q=\rho_{11}-\rho_{22}+\rho_{33}-\rho_{44}$.
For this state, $I^{A\rightarrow B}$ in Eq.(\ref{w5}) gives by
\begin{eqnarray}\label{w8}
I^{A\rightarrow B}&=&\frac{1}{2}[(1+c_3+p+q)\log(1+c_3+p+q) \nonumber\\
&+&(1+c_3-p-q)\log(1+c_3-p-q) \nonumber\\
&+&(1-c_3-p+q)\log(1-c_3-p+q) \nonumber\\
&+& (1-c_3+p-q)\log(1-c_3+p-q)] \nonumber\\
&+&\sum_{i=1,2}[(1+c_i)\log(1+c_i)+(1-c_i)\log(1-c_i)] \nonumber\\
&-& (1+p)\log(1+p)-(1-p)\log(1-p),
\end{eqnarray}
with the base of logarithms being $2$.
The steerability from Bob to Alice can be obtained by exchanging the roles of $A$ and $B$, which is given by
\begin{eqnarray}\label{w9}
S^{B\rightarrow A}={\rm{max}}\bigg\{0, \frac{I^{B\rightarrow A}-2}{I_{\rm{max}}-2}\bigg\},
\end{eqnarray}
with
\begin{eqnarray}\label{w10}
I^{B\rightarrow A}&=&\frac{1}{2}[(1+c_3+p+q)\log(1+c_3+p+q) \nonumber\\
&+&(1+c_3-p-q)\log(1+c_3-p-q) \nonumber\\
&+&(1-c_3-p+q)\log(1-c_3-p+q) \nonumber\\
&+& (1-c_3+p-q)\log(1-c_3+p-q)] \nonumber\\
&+&\sum_{i=1,2}[(1+c_i)\log(1+c_i)+(1-c_i)\log(1-c_i)] \nonumber\\
&-& (1+q)\log(1+q)-(1-q)\log(1-q).
\end{eqnarray}

Unlike quantum entanglement, quantum steering may not be symmetrical, i.e. $S^{A\rightarrow B}\neq S^{B\rightarrow A}$. Based on the asymmetry of quantum steering, we distinguish the
quantum steering into three cases: (i) no-way steering $S^{A\rightarrow B}= S^{B\rightarrow A}=0$; (ii) one-way steering  $S^{A\rightarrow B}>0$ and $ S^{B\rightarrow A}=0$, or vice versa $S^{B\rightarrow A}>0$ and $ S^{A\rightarrow B}=0$; (iii)  two-way steering  $S^{A\rightarrow B}>0$ and $ S^{B\rightarrow A}>0$. This means that, compared with other quantum correlations, quantum steering has richer properties.

\subsection{Quantification of steering based on quantum entanglement }
Firstly, we mention that quantum entanglement
of bipartite states can be effectively identified by the
concurrence. For the X-state $\rho_x$ given by Eq.(\ref{w6}), the concurrence can be expressed as \cite{LLL45}
\begin{eqnarray}\label{A7}
C(\rho_x)=2\max\{|\rho_{14}|-\sqrt{\rho_{22}\rho_{33}}, |\rho_{23}|-\sqrt{\rho_{11}\rho_{44}}\}.
\end{eqnarray}

Next, for any two-qubit state $\rho_{AB}$ shared by Alice and Bob, the steering from Bob to Alice
can be witnessed if the density matrix $\tau_{AB}^1$ defined as \cite{LLL46,LLLL46}
\begin{eqnarray}\label{qq1}
\tau_{AB}^1=\frac{\rho_{AB}}{\sqrt{3}}+\frac{3-\sqrt{3}}{3}(\rho_A\otimes\frac{I}{2}),
\end{eqnarray}
is entangled. Here $\rho_A$ is Alice's reduced density matrix, namely $\rho_A=\rm{Tr}_B(\rho_{AB})$, and  $I$ denotes the two-dimension identity matrix for Bob's qubit.
Similarly, the corresponding steering from
Alice to Bob can be witnessed if
the state $\tau_{AB}^2$ defined as
\begin{eqnarray}\label{qq2}
\tau_{AB}^2=\frac{\rho_{AB}}{\sqrt{3}}+\frac{3-\sqrt{3}}{3}(\frac{I}{2}\otimes\rho_B),
\end{eqnarray}
is entangled, where $\rho_B=\rm{Tr}_A(\rho_{AB})$.

Now, for the X-state $\rho_x$ given by Eq.(\ref{w6}), the matrix $\tau_{AB}^{1}$ can be written as
\begin{eqnarray}\label{qq3}
\tau_{AB}^{1,x}= \left(\!\!\begin{array}{cccc}
\frac{\sqrt{3}}{3}\rho_{11}+r&0&0&\frac{\sqrt{3}}{3}\rho_{14}\\
0&\frac{\sqrt{3}}{3}\rho_{22}+r&\frac{\sqrt{3}}{3}\rho_{23}&0\\
0&\frac{\sqrt{3}}{3}\rho_{23}&\frac{\sqrt{3}}{3}\rho_{33}+s&0\\
\frac{\sqrt{3}}{3}\rho_{14}&0&0&\frac{\sqrt{3}}{3}\rho_{44}+s)
\end{array}\!\!\right),
\end{eqnarray}
with $r=\frac{(3-\sqrt{3})}{6}(\rho_{11}+\rho_{22})$ and $s=\frac{(3-\sqrt{3})}{6}(\rho_{33}+\rho_{44})$. According to Eq.(\ref{A7}), as long as one of the following inequalities,
\begin{eqnarray}\label{qq5}
|\rho_{14}|^2>Q_a-Q_b,
\end{eqnarray}
or
\begin{eqnarray}\label{qqq5}
|\rho_{23}|^2>Q_c-Q_b,
\end{eqnarray}
is satisfied, then the state $\tau_{AB}^{1,x}$ is entangled, where
\begin{eqnarray}
 \nonumber&&Q_a=\frac{2-\sqrt{3}}{2}\rho_{11}\rho_{44}+\frac{2+\sqrt{3}}{2}\rho_{22}\rho_{33}
+\frac{1}{4}(\rho_{11}+\rho_{44})(\rho_{22}+\rho_{33}),\\ \nonumber
&&Q_b=\frac{1}{4}(\rho_{11}-\rho_{44})(\rho_{22}-\rho_{33}),\\ \nonumber
&& Q_c=\frac{2+\sqrt{3}}{2}\rho_{11}\rho_{44}+\frac{2-\sqrt{3}}{2}\rho_{22}\rho_{33}
+\frac{1}{4}(\rho_{11}+\rho_{44})(\rho_{22}+\rho_{33}) \nonumber.
\end{eqnarray}
The steering from  Bob to Alice is thus witnessed.
In a similar way, the steering from Alice to Bob can be witnessed through one of the inequalities,
\begin{eqnarray}\label{qq6}
|\rho_{14}|^2>Q_a+Q_b,
\end{eqnarray}
or
\begin{eqnarray}\label{qqq6}
|\rho_{23}|^2>Q_c+Q_b.
\end{eqnarray}

Further, we introduce the quantities
\begin{eqnarray}\label{qq8}
T^{B\rightarrow A}={\rm{max}}\bigg\{0,\frac{8}{\sqrt{3}}(|\rho_{14}|^2-Q_a+Q_b),
\frac{8}{\sqrt{3}}(|\rho_{23}|^2-Q_c+Q_b)\bigg\},
\end{eqnarray}
and
\begin{eqnarray}\label{qq7}
T^{A\rightarrow B}={\rm{max}}\bigg\{0,\frac{8}{\sqrt{3}}(|\rho_{14}|^2-Q_a-Q_b),
\frac{8}{\sqrt{3}}(|\rho_{23}|^2-Q_c-Q_b)\bigg\},
\end{eqnarray}
to quantify the steerability from Bob to Alice and from Alice to Bob respectively.
The factor $\frac{8}{\sqrt{3}}$ guarantees that the steerability of the maximum entangled state is $1$. Note that we here use the capital letter $T$ to describe the entanglement-based steerability, so as to distinguish from marker $S$ of the steerability  based on entropy uncertainty relation.

\section{Quantization of Dirac fields in Schwarzschild spcetime }

Let's consider a Schwarzschild black hole that is given by the metric \cite{Q42}
\begin{eqnarray}\label{w11}
ds^2&=&-(1-\frac{2M}{r}) dt^2+(1-\frac{2M}{r})^{-1} dr^2\nonumber\\&&+r^2(d\theta^2
+\sin^2\theta d\varphi^2),
\end{eqnarray}
where $r$ and $M$ are respectively the radius and mass of the black hole.
For simplicity, we take $\hbar, G, c$ and $k$ as unity in this paper.
The Dirac equation \cite{Q55} $[\gamma^a e_a{}^\mu(\partial_\mu+\Gamma_\mu)]\Phi=0$ in Schwarzschild spacetime can be written as
\begin{eqnarray}\label{w12}
&&-\frac{\gamma_0}{\sqrt{1-\frac{2M}{r}}}\frac{\partial \Phi}{\partial t}+\gamma_1\sqrt{1-\frac{2M}{r}}\bigg[\frac{\partial}{\partial r}+\frac{1}{r}+\frac{M}{2r(r-2M)} \bigg]\Phi \nonumber\\
&&+\frac{\gamma_2}{r}(\frac{\partial}{\partial \theta}+\frac{\cot \theta}{2})\Phi+\frac{\gamma_3}{r\sin\theta}\frac{\partial\Phi}{\partial\varphi}=0,
\end{eqnarray}
where $\gamma_i$ ($i=0,1,2,3$) are the  Dirac matrices \cite{Q56,Q57}.

Solving the Dirac equation near the event horizon, we obtain a set of positive (fermions) frequency outgoing solutions inside and outside regions of the event horizon \cite{Q56,Q57}
\begin{eqnarray}\label{w13}
\Phi^+_{{\bold k},{\rm in}}\sim \phi(r) e^{i\omega u},
\end{eqnarray}
\begin{eqnarray}\label{w14}
\Phi^+_{{\bold k},{\rm out}}\sim \phi(r) e^{-i\omega u},
\end{eqnarray}
where $\phi(r)$ denotes four-component Dirac spinor, $u=t-r_{*}$ with $r_{*}=r+2M\ln\frac{r-2M}{2M}$ is the tortoise coordinate. $\bold k$ and $\omega$ are the wave vector and frequency which fulfill $|\mathbf{k}|=\omega$ for the massless Dirac field.
The Dirac field $\Phi$ can be expanded as
\begin{eqnarray}\label{w15}
\Phi&=&\int
d\bold k[\hat{a}^{\rm in}_{\bold k}\Phi^{+}_{{\bold k},\text{in}}
+\hat{b}^{\rm in\dag}_{\bold k}
\Phi^{-}_{{\bold k},\text{in}}\nonumber\\ &+&\hat{a}^{\rm out}_{\bold k}\Phi^{+}_{{\bold k},\text{out}}
+\hat{b}^{\rm out\dag}_{\bold k}\Phi^{-}_{{\bold k},\text{out}}],
\end{eqnarray}
where $\hat{a}^{\rm in}_{\bold k}$ and $\hat{b}^{\rm in\dag}_{\bold k}$ are the fermion annihilation and antifermion creation operators for the quantum field in the interior of the event horizon, and $\hat{a}^{\rm out}_{\bold k}$ and $\hat{b}^{\rm out\dag}_{\bold k}$ are the fermion annihilation and antifermion creation operators for the quantum field of the
exterior region, respectively. These annihilation and creation operators satisfy canonical anticommutation $\{\hat{a}^{\rm out}_{\mathbf{k}},\hat{a}^{\rm out\dagger}_{\mathbf{k'}}\}=
\{\hat{b}^{\rm in}_{\mathbf{k}},\hat{b}^{\rm in\dagger}_{\mathbf{k'}}\}
=\delta_{\mathbf{k}\mathbf{k'}}. $ One  can  define the Schwarzschild vacuum through expression $\hat{a}^{\rm in}_{\bold k}|0\rangle_S=\hat{a}^{\rm out}_{\bold k}|0\rangle_S=0$.
Therefore, the
modes  $\Phi^\pm_{{\bold k},{\rm in}}$ and $\Phi^\pm_{{\bold k},{\rm out}}$ are usually called Schwarzschild modes.

Making an analytic continuation of Eqs.(\ref{w13}) and (\ref{w14}) in the light of Domour and Ruffini's suggestion \cite{Q58}, one
find a complete basis for positive energy modes, i.e., the
Kruskal modes,.
\begin{eqnarray}\label{w16}
\Psi^+_{{\bold k},{\rm out}}=e^{-2\pi M\omega} \Phi^-_{{-\bold k},{\rm in}}+e^{2\pi M\omega}\Phi^+_{{\bold k},{\rm out}},
\end{eqnarray}
\begin{eqnarray}\label{w17}
\Psi^+_{{\bold k},{\rm in}}=e^{-2\pi M\omega} \Phi^-_{{-\bold k},{\rm out}}+e^{2\pi M\omega}\Phi^+_{{\bold k},{\rm in}}.
\end{eqnarray}
Thus, we can also use the Kruskal modes to expand the Dirac fields in the Kruskal spacetime
\begin{eqnarray}\label{w18}
\Phi&=&\int
d\bold k [2\cosh(4\pi M\omega)]^{-\frac{1}{2}}
[\hat{c}^{\rm in}_{\bold k}\Psi^{+}_{{\bold k},\text{in}}
+\hat{d}^{\rm in\dag}_{\bold k}
\Psi^{-}_{{\bold k},\text{in}}\nonumber\\ &+&\hat{c}^{\rm out}_{\bold k}\Psi^{+}_{{\bold k},\text{out}}
+\hat{d}^{\rm out\dag}_{\bold k}\Psi^{-}_{{\bold k},\text{out}}],
\end{eqnarray}
where $\hat{c}^{\sigma}_{\bold k}$ and $\hat{d}^{\sigma\dag}_{\bold k}$ with $\sigma=(\rm in, \rm out)$ are the fermion annihilation and antifermion creation operators acting on the Kruskal vacuum.

Eqs.(\ref{w15}) and (\ref{w18}) represent the different decompositions of the same Dirac field in Schwarzschild and  Kruskal modes, respectively, which lead to the well-known Bogoliubov transformation between the Kruskal and Schwarzschild
operators,
\begin{eqnarray}\label{w19}
\hat{c}^{\rm out}_{\bold k}&=&\frac{1}{\sqrt{e^{-8\pi M\omega}+1}}\hat{a}^{\rm out}_{\bold k}-\frac{1}{\sqrt{e^{8\pi M\omega}+1}}\hat{b}^{\rm out\dag}_{\bold k},\\
\hat{c}^{\rm out\dag}_{\bold k}&=&\frac{1}{\sqrt{e^{-8\pi M\omega}+1}}\hat{a}^{\rm out\dag}_{\bold k}-\frac{1}{\sqrt{e^{8\pi M\omega}+1}}\hat{b}^{\rm out}_{\bold k}.
\end{eqnarray}
The Kruskal vacuum and excited states thus can be expressed in the Schwarzschild Fock space as
\begin{eqnarray}\label{w20}
\nonumber |0\rangle_K&=&\frac{1}{\sqrt{e^{-\frac{\omega}{T}}+1}}|0\rangle_{\rm out} |0\rangle_{\rm in}+\frac{1}{\sqrt{e^{\frac{\omega}{T}}+1}}|1\rangle_{\rm out} |1\rangle_{\rm in},\\
|1\rangle_K&=&|1\rangle_{\rm out} |0\rangle_{\rm in},
\end{eqnarray}
where $T=\frac{1}{8\pi M}$ is the Hawking temperature, $\{|n\rangle_{\rm out}\}$ and $\{|n\rangle_{\rm in}\}$ are the Schwarzschild number states for the fermion outside the region and the antifermion inside the region of the event horizon, respectively.

For the Schwarzschild observer Bob who  hovers outside the event horizon, the
Hawking radiation spectrum from the viewpoint of his detector
is given by \cite{Q57}
\begin{eqnarray}\label{w21}
N_F^2=\sideset{_K}{}{\mathop{\langle}}0|\hat{a}^{\rm out\dag}_{\bold k}\hat{a}^{\rm out}_{\bold k}|0\rangle_K=\frac{1}{e^{\frac{\omega}{T}}+1}.
\end{eqnarray}
This equation means that a Kruskal vacuum observed by the Kruskal observer Alice would be detected as a number of the generated fermions $N_F^2$  from Bob's viewpoint.
In other words, the Schwarzschild observer Bob in the exterior of
the black hole can detect a thermal Fermi-Dirac statistic of fermions.

\section{Fermionic steering in Schwarzschild spacetime}
Consider two maximally entangled fermionic modes in the asymptotically flat region of the
Schwarzschild black hole
\begin{eqnarray}\label{w22}
|\phi_{AB}\rangle=\frac{1}{\sqrt{2}}(|0\rangle_A|0\rangle_B+|1\rangle_A|1\rangle_B),
\end{eqnarray}
where the subscripts $A$ and $B$ denote the modes which are associated
with the observers Alice and Bob, respectively.
After the coincidence of Alice and Bob, Alice stays stationary at the
asymptotically flat region, while Bob hovers outside the event horizon
of the black hole. Bob will detects a thermal Fermi-Dirac statistic of fermions and his detector is found to be excited.
Using Eq.(\ref{w20}), we can rewrite Eq.(\ref{w22}) in terms of Kruskal  modes for Alice
and Schwarzschild modes for Bob
\begin{eqnarray}\label{w23}
|\phi_{AB\bar B}\rangle&=&\frac{1}{\sqrt{2}}(\frac{1}{\sqrt{e^{-\frac{\omega}{T}}+1}}|0\rangle_{ A}|0\rangle_B|0\rangle_{\bar B}+\frac{1}{\sqrt{e^{\frac{\omega}{T}}+1}}|0\rangle_A|1\rangle_B|1\rangle_{\bar B} \nonumber\\
&+&|1\rangle_A|1\rangle_B|0\rangle_{\bar B}),
\end{eqnarray}
where the mode $\bar B$ is observed by a hypothetical observer Anti-Bob inside
the event horizon of the black hole. We write its density matrix as
\begin{eqnarray}\label{w24}
 \rho_{AB\bar B}=\frac{1}{2}\left(\!\!\begin{array}{cccccccc}
\mathcal{C}^{2}&0&0&\mathcal{C}\mathcal{S}&0&0&\mathcal{C}&0\\
0&0&0&0&0&0&0&0\\
0&0&0&0&0&0&0&0\\
\mathcal{C}\mathcal{S}&0&0&\mathcal{S}^{2}&0&0&\mathcal{S}&0\\
0&0&0&0&0&0&0&0\\
0&0&0&0&0&0&0&0\\
\mathcal{C}&0&0&\mathcal{S}&0&0&1&0\\
0&0&0&0&0&0&0&0
\end{array}\!\!\right).
\end{eqnarray}
in the orthonormal basis  $\{|0,0,0\rangle,|0,0,1\rangle,|0,1,0\rangle,|0,1,1\rangle,|1,0,0\rangle,|1,0,1\rangle,|1,1,0\rangle,|1,1,1\rangle\}$,
where we have defined $|abc\rangle=|a\rangle_A|b\rangle_B|c\rangle_{\bar B}$, and $\mathcal{C}=\frac{1}{\sqrt{e^{-\frac{\omega}{T}}+1}}$, $\mathcal{S}=\frac{1}{\sqrt{e^{\frac{\omega}{T}}+1}}$ for simplicity.

\subsection{ Physically accessible quantum steering }
Since Bob is causally disconnected from the region inside the event horizon, the
only information that is physically accessible to the observers
is encoded in the mode $A$ described by Alice and the mode $B$ outside the event horizon
described by Bob. Taking the trace over the $\bar B$ mode inside the event horizon,
we obtain a mixed density matrix for Alice and Bob
\begin{eqnarray}\label{w25}
\rho_{AB}=\frac{1}{2} \left(\!\!\begin{array}{cccccccc}
\mathcal{C}^2 & 0 & 0 & \mathcal{C} \\
0 & \mathcal{S}^2 &0 &0 \\
0 & 0 & 0 & 0\\
\mathcal{C} & 0 & 0 & 1
\end{array}\!\!\right),
\end{eqnarray}
in the basis $\{|00\rangle,|01\rangle,|10\rangle,|11\rangle\}$. In the following, we use  two types of methods to measure steering in curved spacetime.

Firstly, according to Eqs.(\ref{w5}) and (\ref{w9}), we obtain the analytic
expressions for the steerability of the $S^{A\rightarrow B}$ and $S^{B\rightarrow A}$ based on entropy uncertainty relation as
\begin{eqnarray}\label{w26}
S^{A\rightarrow B}(\rho_{AB})&=&\frac{1}{4}\big[2(1+\mathcal{C})\log(1+\mathcal{C})
+2(1-\mathcal{C})\log(1-\mathcal{C})\nonumber\\
&+&\mathcal{C}^2\log(\mathcal{C}^2)
+\mathcal{S}^2\log(\mathcal{S}^2)\big],
\end{eqnarray}
and
\begin{eqnarray}\label{w28}
S^{B\rightarrow A}(\rho_{AB})&=&\frac{1}{4}\big[2(1+\mathcal{C})\log(1+\mathcal{C})
+2(1-\mathcal{C})\log(1-\mathcal{C})\nonumber\\
&-&(1+\mathcal{S}^2)\log(1+\mathcal{S}^2)+\mathcal{S}^2\log(\mathcal{S}^2)].
\end{eqnarray}
We see that the steerability depends on the Hawking temperature $T$, i.e., Hawking radiation of the black hole influences the fermionic steerability. Under the influence of Hawking radiation, the steering becomes asymmetric, i.e., the steerability from Alice to Bob is different from the steerability from Bob to Alice. In order to measure the degrees of asymmetry, we introduce the steering difference
\begin{eqnarray}\label{w27}
S_{AB}^\Delta&=&|S^{A\rightarrow B}(\rho_{AB})-S^{B\rightarrow A}(\rho_{AB})|\\
&=&\frac{1}{4}[\mathcal{C}^2\log(\mathcal{C}^2)+
(1+\mathcal{S}^2)\log(1+\mathcal{S}^2)] \nonumber.
\end{eqnarray}

Secondly, we study another quantification of steering based on quantum entanglement in Schwarzschild spacetime.  The analytic expressions for the entanglement-based steerability $T^{A\rightarrow B}$, $ T^{B\rightarrow A}$ and the corresponding steering difference read
\begin{eqnarray}\label{qq10}
T^{A\rightarrow B}(\rho_{AB})&=&\mathcal{C}^2-\frac{1}{\sqrt{3}}\mathcal{C}^2\mathcal{S}^2,
\end{eqnarray}
\begin{eqnarray}\label{qq11}
T^{B\rightarrow A}(\rho_{AB})&=&\mathcal{C}^2-\frac{1}{\sqrt{3}}\mathcal{S}^2,
\end{eqnarray}
and
\begin{eqnarray}\label{qq12}
T_{AB}^\Delta=|T^{A\rightarrow B}(\rho_{AB})-T^{B\rightarrow A}(\rho_{AB})|
=\frac{1}{\sqrt{3}}\mathcal{S}^4,
\end{eqnarray}
respectively.

\begin{figure}
\begin{minipage}[t]{0.5\linewidth}
\centering
\includegraphics[width=3.0in,height=5.2cm]{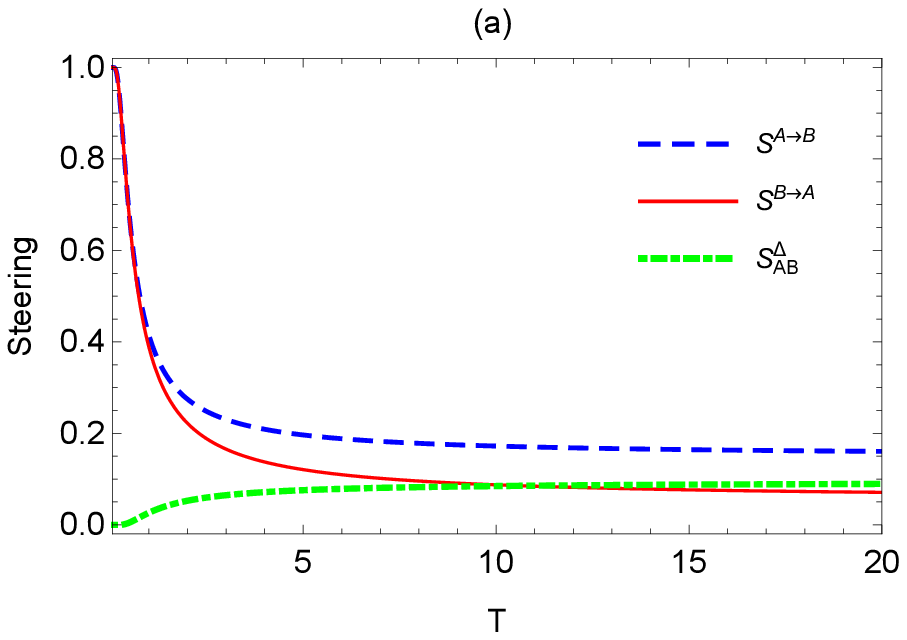}
\label{fig1a}
\end{minipage}%
\begin{minipage}[t]{0.5\linewidth}
\centering
\includegraphics[width=3.0in,height=5.2cm]{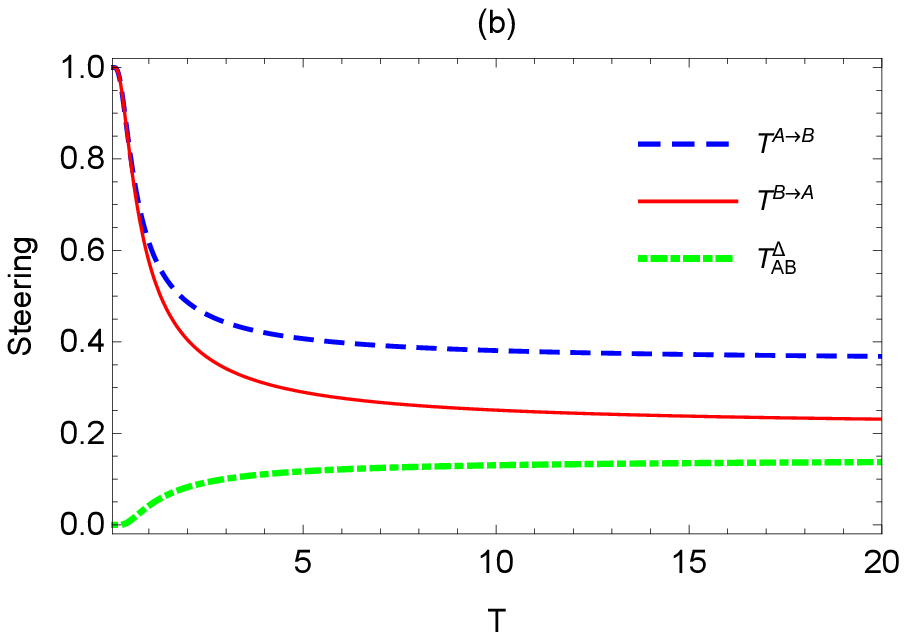}
\label{fig1c}
\end{minipage}%
\caption{The fermionic steerability and steering asymmetry between Alice and Bob as functions of the Hawking temperature $T$. The
parameter $\omega$ is fixed as $\omega=1$.}
\label{Fig1}
\end{figure}

In Fig.\ref{Fig1},  we plot the $A\rightarrow B$
steering, $B\rightarrow A$ steering  and  steering asymmetry as functions of the
Hawking temperature $T$. We find that the results for the two types of measures are consistent. With the increase of the Hawking temperature $T$, the fermionic steerability between Alice and Bob decreases firstly and then approaches to the nonzero asymptotic value in the infinite Hawking temperature. The $A\rightarrow B$ fermionic steerability is always bigger than
the $B\rightarrow A$  fermionic  steerability. The steering asymmetry increases with Hawking temperature and approaches to the asymptotic values,
$$ \lim_{T\rightarrow\infty}S_{AB}^\Delta\approx0.0944,\hspace{1cm} \lim_{T\rightarrow\infty}T_{AB}^\Delta=\frac{\sqrt{3}}{12},$$
for infinite Hawking temperature. These results contrast sharply with case of bosonic fields \cite{Q53,Q54,Q59}. With the increase of Hawking temperature, the bosonic steerability reduces quickly and suffers from a``sudden death". The $A\rightarrow B$ bosonic steerability is always smaller than the $B\rightarrow A$ bosonic steerability, and the bosonic steering asymmetry increases firstly and then decreases to zero when Hawking temperature increases.
These clearly different results between bosonic steerability and fermionic  steerability originate from the difference between the Fermi-Dirac statistic and the Bose-Einstein statistic, which perhaps are available in practice.
For example, if we need the steerability from Alice to Bob over the steerability from Bob to Alice in curved spacetime, then we should use fermionic steering rather than bosonic steering.

\subsection{ Physically inaccessible quantum steering }

Besides the steering between Alice and Bob, we can also discuss the steering between Alice and Anti-Bob, and the steering between Bob and Anti-Bob. As Anti-Bob is inside of the event horizon, we use the term ``inaccessible steering".

(i) The fermionic steering  between Alice and Anti-Bob.
Tracing over the mode $B$ held by Bob, we obtain the density matrix
for subsystem Alice and Anti-Bob
\begin{eqnarray}\label{w29}
\rho_{A\bar B}=\frac{1}{2} \left(\!\!\begin{array}{cccccccc}
\mathcal{C}^2 & 0 & 0 & 0 \\
0 & \mathcal{S}^2 &\mathcal{S} &0 \\
0 & \mathcal{S} & 1 & 0\\
0 & 0 & 0 & 0
\end{array}\!\!\right).
\end{eqnarray}
Following the calculation steps in above subsection, we obtain the fermionic steering between Alice and Anti-Bob, as well as the corresponding steering asymmetry for the two types of quantifications as
\begin{eqnarray}\label{w30}
S^{A\rightarrow \bar B}(\rho_{A\bar B})&=&\max\bigg\{0,\frac{1}{4}\big[2(1+\mathcal{S})\log(1+\mathcal{S})
+2(1-\mathcal{S})\log(1-\mathcal{S})\nonumber\\
&+&\mathcal{C}^2\log(\mathcal{C}^2)
+\mathcal{S}^2\log(\mathcal{S}^2)\big]\bigg\},
\end{eqnarray}
\begin{eqnarray}\label{w31}
S^{\bar{B}\rightarrow A}(\rho_{AB})&=&\max\bigg\{0,\frac{1}{4}\big[2(1+\mathcal{S})\log(1+\mathcal{S})+2(1-\mathcal{S})\log(1-\mathcal{S})\nonumber\\
&-&(1+\mathcal{C}^2)\log(1+\mathcal{C}^2)
+\mathcal{C}^2\log(\mathcal{C}^2)\big]\bigg\},
\end{eqnarray}
\begin{eqnarray}\label{w31}
S_{A\bar B}^\Delta=|S^{A\rightarrow \bar B}(\rho_{A\bar B})-S^{\bar B\rightarrow A}(\rho_{A\bar B})|,
\end{eqnarray}
and
\begin{eqnarray}\label{qq13}
T^{A\rightarrow \bar B}(\rho_{A\bar B})=\max\bigg\{0,\mathcal{S}^2(1-\frac{1}{\sqrt{3}}\mathcal{C}^2)\bigg\},
\end{eqnarray}
\begin{eqnarray}\label{qq14}
T^{\bar B\rightarrow A}(\rho_{A\bar B})=\max\bigg\{0,\mathcal{S}^2-\frac{1}{\sqrt{3}}\mathcal{C}^2\bigg\}
\end{eqnarray}
\begin{eqnarray}\label{qqq15}
T_{A\bar B}^\Delta=|T^{A\rightarrow \bar B}(\rho_{A\bar B})-T^{\bar B\rightarrow A}(\rho_{A\bar B})|.
\end{eqnarray}

\begin{figure}
\begin{minipage}[t]{0.5\linewidth}
\centering
\includegraphics[width=3.0in,height=5.2cm]{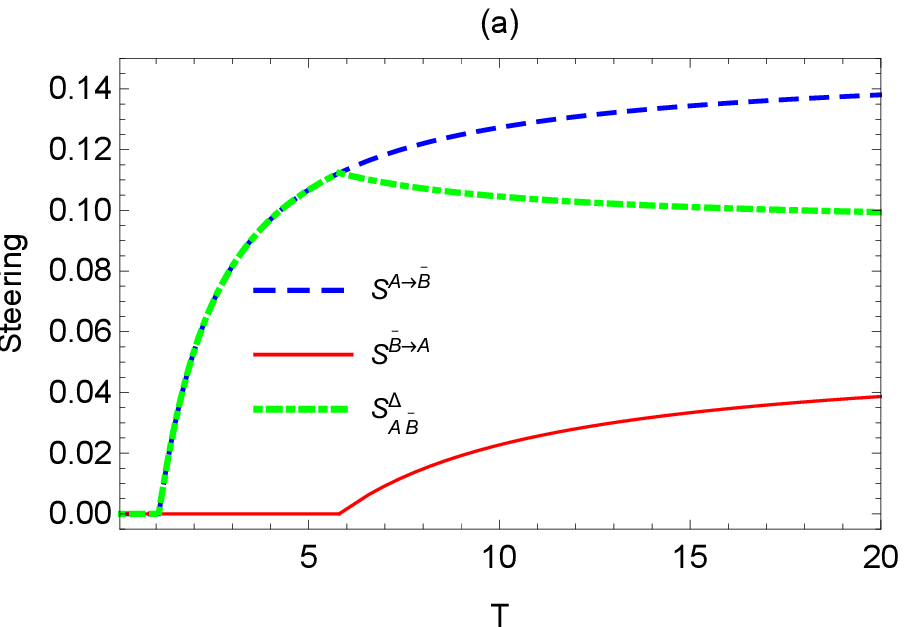}
\label{fig1a}
\end{minipage}%
\begin{minipage}[t]{0.5\linewidth}
\centering
\includegraphics[width=3.0in,height=5.2cm]{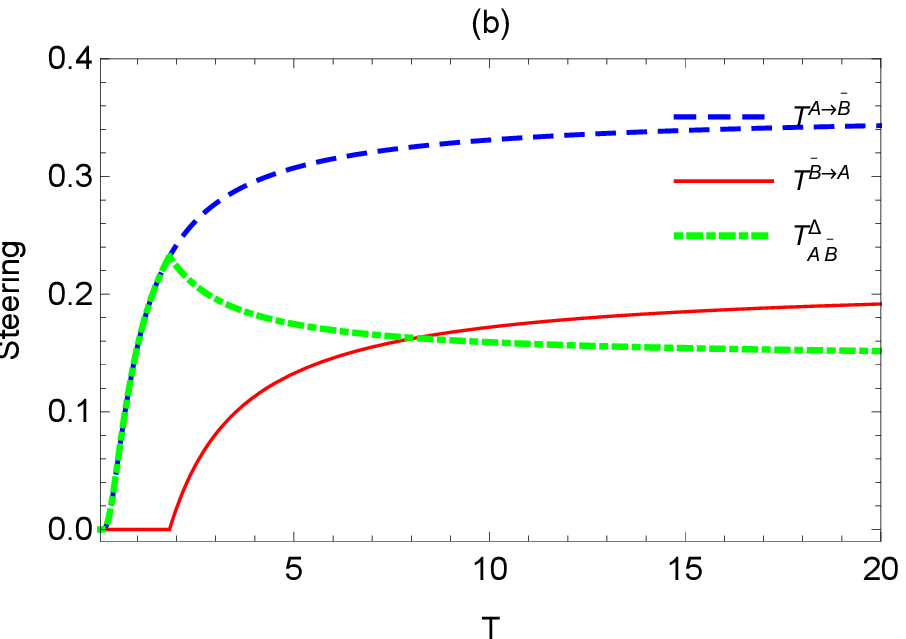}
\label{fig1c}
\end{minipage}%
\caption{The fermionic steerability and steering asymmetry  between Alice and Anti-Bob as functions of the
Hawking temperature $T$. The
parameter $\omega$ is fixed as $\omega=1$. }
\label{Fig2}
\end{figure}

Fig.\ref{Fig2} shows how the Hawking temperature $T$ of the black hole
influences the fermionic steerability and steering asymmetry  between Alice and Anti-Bob.
We see that results for the two type of quantifications are basically similar. Hawking radiation can generate fermionic steering between Alice and Anti-Bob, while the temperature for generating $A\rightarrow \bar B$ steering is always lower than the temperature for generating $\bar B \rightarrow A$ steering. As the temperatures and speeds for generating $A\rightarrow \bar B$ steering and $\bar B\rightarrow  A$ steering are different, the steering asymmetry appears. The steering asymmetry firstly increases to the maximum and then decreases to the nonzero asymptotic value with the increase of
the Hawking temperature $T$. The maximal steering asymmetry takes place at the temperature where the $\bar B\rightarrow  A$ steering births, i.e., takes place at the transition point from one-way steering to two-way steering.
For the parameters taken in the figure, the Hawking temperatures for the maximal steering asymmetry are $T_s\approx5.8021\omega$ [Fig.2(a)] and $T_t=\frac{\omega}{\ln\sqrt{3}}$ [Fig.2(b)], respectively. The figure also shows that the steerability between Alice and Anti-Bob approaches to the finite asymptotic value when $T\rightarrow \infty$, which fulfills the intriguing relations,
$$\lim_{T\rightarrow\infty}S^{A\rightarrow \bar B}=\lim_{T\rightarrow\infty}S^{A\rightarrow B}, \lim_{T\rightarrow\infty}S^{\bar B\rightarrow A}=\lim_{T\rightarrow\infty}S^{B\rightarrow A}, $$
$$\lim_{T\rightarrow\infty}T^{A\rightarrow \bar B}=\lim_{T\rightarrow\infty}T^{A\rightarrow B}, \lim_{T\rightarrow\infty}T^{\bar B\rightarrow A}=\lim_{T\rightarrow\infty}T^{B\rightarrow A}. $$
This means that the steering status for Bob and Anti-Bob are the same when $T\rightarrow \infty$.

Comparing Fig.2(a) and (b), we find that the two types of quantifications for steering also have some tiny difference. For quantification of steering based on entropy uncertainty relation, both $S^{A\rightarrow \bar B}$ and  $S^{\bar B\rightarrow A}$ appear ``sudden birth" behavior with the growth of the Hawking temperature. For the quantification of steering on quantum entanglement, however, only the steerability $T^{\bar B\rightarrow A}$ appear as ``sudden birth". Further, the temperature for generating two-way steering for entanglement-based quantification is lower than the case for the quantification based on entropy uncertainty relation. In this sense, we can say that the quantification of steering based on entanglement is more sensitive than the quantification of steering based on entropy uncertainty relation.

It has been shown that Hawking radiation for bosonic fields cannot generate entanglement and steering between Alice and Anti-Bob\cite{Q40,Q54,QWE59}, which makes sharp contrast with the fermionic steering discussed above.
We may understand the reason as follows: The Hawking radiation for bosonic fields is equivalent to a local operation on the subsystem of Bob and Anti-Bob, which cannot generate bosonic entanglement or steering
between Alice and Anti-Bob. For the fermionic fields, however, the Pauli exclusion principle inhibits the production of more than one fermion in one mode. This limitation forms an interaction between Alice and Bob (or Ant-Bob). Therefore, the entanglement (or steering) between Alice and Anti-Bob can be produced \cite{Q60}.

(ii) The fermionic steering between Bob and Anti-Bob. Tracing
over the mode $A$, we obtain the density matrix for subsystem of Bob and Anti-Bob
\begin{eqnarray}\label{w32}
\rho_{B\bar B}= \frac{1}{2}\left(\!\!\begin{array}{cccccccc}
\mathcal{C}^2 & 0 & 0 & \mathcal{C}\mathcal{S} \\
0 & 0 &0 &0 \\
0 & 0 & 1 & 0\\
\mathcal{C}\mathcal{S} & 0 & 0 & \mathcal{S}^2
\end{array}\!\!\right).
\end{eqnarray}
The fermionic steering between Bob and Anti-Bob are thus calculated as
\begin{eqnarray}\label{w34}
S^{B\rightarrow \bar B}(\rho_{A\bar B})&=&\max\bigg\{0,\frac{1}{4}\big[2(1+\mathcal{C}\mathcal{S})\log(1+\mathcal{C}\mathcal{S})
+2(1-\mathcal{C}\mathcal{S})\log(1-\mathcal{C}\mathcal{S})\nonumber\\
&-&(1+\mathcal{S}^2)\log(1+\mathcal{S}^2)
+\mathcal{S}^2\log(\mathcal{S}^2)\big]\bigg\}=0,
\end{eqnarray}
\begin{eqnarray}\label{w35}
S^{\bar B\rightarrow B}(\rho_{A\bar B})&=&\max\bigg\{0,\frac{1}{4}\big[2(1+\mathcal{C}\mathcal{S})\log(1+\mathcal{C}\mathcal{S})
+2(1-\mathcal{C}\mathcal{S})\log(1-\mathcal{C}\mathcal{S})\nonumber\\
&-&(1+\mathcal{C}^2)\log(1+\mathcal{C}^2)
+\mathcal{C}^2\log(\mathcal{C}^2)\big]\bigg\}=0,
\end{eqnarray}
and
\begin{eqnarray}\label{qq15}
T^{B\rightarrow \bar B}(\rho_{B\bar B})=\max\bigg\{0,\mathcal{S}^2\big(\mathcal{C}^2-\frac{1}{\sqrt{3}}\big)\bigg\},
\end{eqnarray}
\begin{eqnarray}\label{qq16}
T^{\bar B\rightarrow B}(\rho_{B\bar B})=\max\bigg\{0,\mathcal{C}^2\big(\mathcal{S}^2-\frac{1}{\sqrt{3}}\big)\bigg\}=0.
\end{eqnarray}
From Eqs.(\ref{w34})-(\ref{qq16}) and combined with Fig.3, we find that Hawking radiation cannot produce fermionic steering based on entropy uncertainty relation between Bob and Anti-Bob, but can produce the entanglement-based steering $T^{B\rightarrow \bar B}$. It again suggests that the quantification of steering based on entanglement is a more sensitive than that based on entropy uncertainty relation.

\begin{figure}
\begin{minipage}[t]{0.5\linewidth}
\centering
\includegraphics[width=3.0in,height=5.2cm]{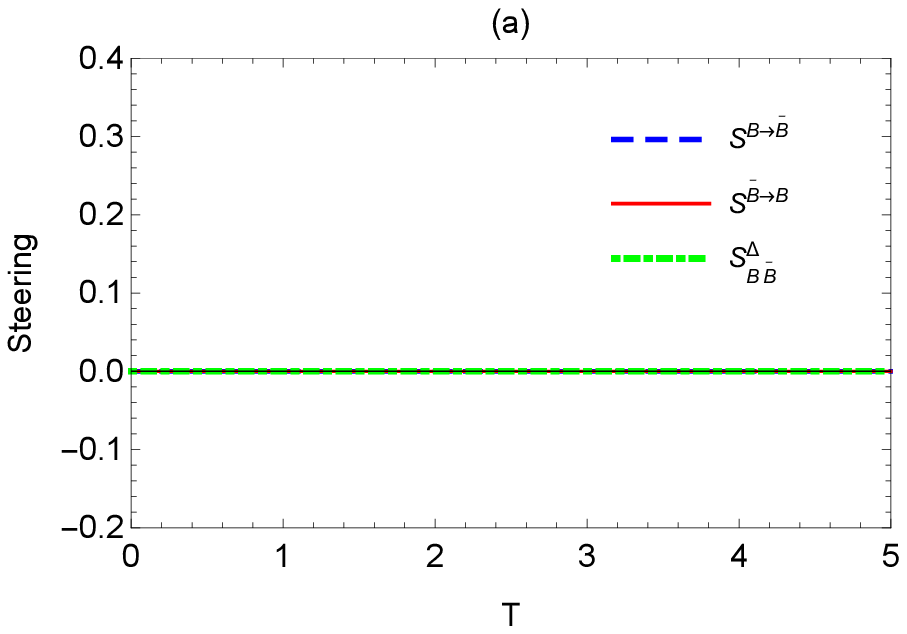}
\label{fig1a}
\end{minipage}%
\begin{minipage}[t]{0.5\linewidth}
\centering
\includegraphics[width=3.0in,height=5.2cm]{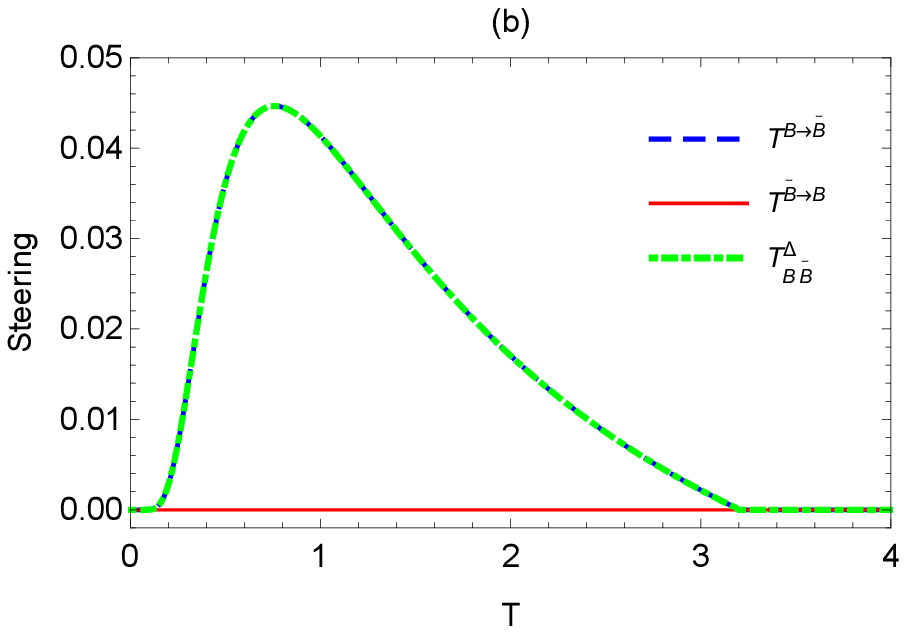}
\label{fig1c}
\end{minipage}%
\caption{The fermionic steerability and steering asymmetry between Bob and Anti-Bob as functions of the Hawking temperature $T$. The
parameter $\omega$ is fixed as $\omega=1$. }
\label{Fig3}
\end{figure}

Fig.\ref{Fig3}(b) shows that the fermionic steering $T^{B\rightarrow \bar B}$
increases from zero to the maximum and then decreases to zero again (sudden death) with the increasing of Hawking temperature.
The Hawking temperature for the maximal steering is $T_{max}=\frac{\omega}{\ln(\sqrt{3}+1)-\ln(\sqrt{3}-1)}$, and for sudden death is $T_{dea}=\frac{-\omega}{\ln(\sqrt{3}-1)}$.
This result is quite different from the behavior of fermionic entanglement between Bob to Anti-Bob in a relativistic setting, where fermionic entanglement increases monotonically \cite{Q60}.

\section{ Monogamous relation between fermionic steering and entanglement in Schwarzschild spacetime}
It is well known that quantum steering is an intermediate
form of quantum inseparabilities in between Bell nonlocality and quantum entanglement \cite{Q3}.
Quantum states that show Bell nonlocality form a strict subset of quantum states
showing quantum steering, the latter also form a strict subset of entangled states.
Quantum steering is a good potential object for connecting  Bell nonlocality and quantum entanglement. In this section, we try to establish the relations between fermionic steering and entanglement
in Schwarzschild spacetime.

From above study, we find that with the increase of the  Hawking temperature $T$, the physically accessible fermionic steering between Alice and Bob decreases monotonically, and at the same time the physically inaccessible fermionic steering between Alice (Bob) and Anti-Bob increases monotonically or nonmonotonically. Also note that the entanglement has similar properties: The physically accessible fermionic entanglement decreases monotonically and the physically inaccessible fermionic entanglement increases monotonically with the growth of the Hawking temperature \cite{Q60}.
A question arises naturally: Are there relations between the physically accessible steering (entanglement) and the physically inaccessible steering (entanglement)?
The answer is yes. Through analytical calculation, we find some relations of this types, which can help us to understand more deeply the fermionic entanglement and steering in a relativistic setting.

From Eqs.(\ref{w25}), (\ref{w29}) and (\ref{w32}), we can calculate the concurrence of entanglement $C(\rho_{A B})$ between Alice and Bob, the concurrence $C(\rho_{A \bar B})$ between Alice and Anti-Bob, as well as the concurrence $C(\rho_{B \bar B})$ between Bob and Anti-Bob as,
$$C(\rho_{A B})=\frac{1}{\sqrt{e^{-\frac{\omega}{T}}+1}},$$ $$C(\rho_{A \bar B})=\frac{1}{\sqrt{e^{\frac{\omega}{T}}+1}},$$ and $$C(\rho_{B \bar B})=\frac{1}{\sqrt{e^{\frac{\omega}{T}}+e^{-\frac{\omega}{T}}+2}}.$$
After carefully inspection, we find the following monogamy relations between fermionic steering and entanglement,
\begin{eqnarray}\label{qq20}
T^{A\rightarrow B}(\rho_{AB})-T^{A\rightarrow \bar B}(\rho_{A\bar B})=C^2(\rho_{A B})-C^2(\rho_{A\bar B}),
\end{eqnarray}
\begin{eqnarray}\label{qq21}
T^{A\rightarrow B}(\rho_{AB})+T^{A\rightarrow \bar B}(\rho_{A\bar B})=C^2(\rho_{A B})+C^2(\rho_{A\bar B})-\frac{2}{\sqrt{3}}C^2(\rho_{B\bar B}),
\end{eqnarray}
\begin{eqnarray}\label{qq22}
\frac{3-\sqrt{3}}{2}[T^{B\rightarrow A}(\rho_{AB})-T^{\bar B\rightarrow A}(\rho_{A\bar B})]=C^2(\rho_{A B})-C^2(\rho_{A\bar B}), ~{\rm for} &\frac{\omega}{\ln\sqrt{3}}<T,
\end{eqnarray}
\begin{eqnarray}\label{qq23}
\frac{3+\sqrt{3}}{2}[T^{B\rightarrow A}(\rho_{AB})+T^{\bar B\rightarrow A}(\rho_{A\bar B})]=C^2(\rho_{A B})+C^2(\rho_{A\bar B}), ~{\rm for} &\frac{\omega}{\ln\sqrt{3}}<T.
\end{eqnarray}
These monogamies reveal the relation between the physically accessible correlation and the physically inaccessible correlation, as well as the relation between entanglement and steering. They suggest that the Hawking radiation can give rise to the transformation between these different types of quantum correlations.

\section{Conclusions}
In conclusion, we have investigated the influence of Hawking radiation on the fermionic quantum steering for the setup where Alice resides in the asymptotically flat region and Bob hovers near the event horizon of a Schwarzschild black hole. Two different types of quantification for quantum steering have been employed. The redistribution and transformation between physically accessible and inaccessible steering induced by Hawking radiation have been studied. Some monogamy relations between fermionic steering and entanglement have been found.

Firstly, we have found that Hawking radiation reduces the fermionic steerability between Alice and Bob, and make it approaching to the nonzero asymptotic values in the limit of infinite Hawking temperature.
The steerability from Alice to Bob is always bigger than the steerability from Bob to Alice in the degradation process. The steering asymmetry increases with Hawking temperature and approaches to the nonzero asymptotic values for infinite Hawking temperature. For the degradation of steering between Alice and Bob, the two types of quantification for steerability behave completely consistent.

Secondly, we have found that Hawking radiation can produce fermionic steering between Alice and Anti-Bob, which approaches to the finite asymptotic values in the limit of infinite Hawking temperature. The temperature for generating $A\rightarrow \bar B$ steering is lower than for generating $\bar B \rightarrow A$ steering. Also, the $A\rightarrow \bar B$ steerability is always greater than $\bar B \rightarrow A$ steerability. The steering asymmetry firstly increases, then decreases and finally approaches to a nonzero asymptotic value when the Hawking temperature changes from zero to infinite. The maximal steering asymmetry occurs at the transition point from one-way steering to two-way steering. Note that here the two types of quantification for steering behave slightly different: The temperature for generating steering based on entanglement is lower than for generating steering based on entropy uncertainty relation, meaning that the quantification of steering based on entanglement is more sensitive than that based on entropy uncertainty relation.

Thirdly, we have also studied the effect of Hawking radiation on the steering between Bob and Anti-Bob. We have found that the two types of quantification for steering in this case behave clearly different. The steering between Bob and Anti-Bob based on entropy uncertainty relation is always zero, but the steering from Bob to Anti-Bob based on entanglement can be generated for some domain of Hawking temperature. This again suggests that the quantification of steering based on entanglement is more sensitive than that based on entropy uncertainty relation.
On the other hand, quantum steering between
Bob hovering near the event horizon and anti-Bob segregated by the event horizon
is different from the other two forms of quantum steering,  possibly because Alice stays stationary at an asymptotically flat region.

Finally, we have established some monogamy relations between fermionic steering and fermionic entanglement. These monogamies reveal the regularity for the redistribution of steering and entanglement between different subsystems under the Hawking effect, and may be useful for understanding the information paradox of black holes.

We have also made a comparison between fermionic steering studied in this context and bosonic steering studied previously under the influence of Hawking radiation. The previous study suggested that the bosonic steering between Alice and Bob suffers from sudden death in the process of degradation, and the $A\rightarrow B$ bosonic steerability is always smaller than the $B\rightarrow A$ bosonic steerability under the Hawking effect \cite{Q53,Q54}. Hawking effect cannot generate the bosonic steering between Alice and Anti-Bob \cite{Q40,Q54,QWE59}. All these results are opposite to the fermionic steering displayed in the text.

\begin{acknowledgments}
This work is supported by the National Natural
Science Foundation of China (Grant Nos.1217050862, 11275064), and 2021BSL013.	
\end{acknowledgments}


\end{document}